\documentclass[oldversion,letter]{aa}
\usepackage{natbib}
\bibpunct{(}{)}{;}{a}{}{,}
\usepackage{graphicx}
\usepackage{eufrak}
\usepackage{times}
\usepackage{txfonts}

\begin{document}

\title{Intense Mass Loss from C-rich AGB Stars at low Metallicity?}
\titlerunning{Models of mass loss from C-rich AGB stars at different metallicities}

\author{L. Mattsson\thanks{\email{mattsson@astro.uu.se}}
        \and R. Wahlin
        \and S. H\"ofner
	  \and K. Eriksson}
\institute{Dept. Physics and Astronomy, Div. of Astronomy and Space Physics, Uppsala University, Box 515, SE-751 20 Uppsala, Sweden}

\offprints{Lars Mattsson}

\date{Received date; accepted date}

\abstract{We argue that the energy injection of pulsations may be of greater importance to the mass-loss rate 
  of AGB stars than metallicity, and that 
  the mass-loss trend with metallicity is not as simple as sometimes assumed. Using our detailed radiation hydrodynamical 
  models that include dust formation, we illustrate the effects of pulsation energy on wind properties.  We find that the 
  mass-loss rate scales with the kinetic energy input by pulsations as long as a dust-saturated wind does not occur,
  and all other stellar parameters are kept constant. This includes the absolute abundance of condensible carbon (not bound in CO),
  which is more relevant than keeping the ${\rm C/O}$-ratio constant when comparing stars of different metallicity. The pressure 
  and temperature gradients in the atmospheres of stars, become steeper and flatter, respectively, when the metallicity is reduced, 
  while the radius where the atmosphere becomes opaque is typically associated with a higher gas pressure. 
  This effect can be compensated for by adjusting the velocity amplitude of the variable inner boundary (piston), which is used
  to simulate the effects of pulsation, to obtain models with comparable kinetic-energy input. Hence, it is more relevant to 
  compare models with similar energy-injections than of similar velocity amplitude. Since there is no evidence for 
  weaker pulsations in low-metallicity AGB stars, we conclude that it is \textit{unlikely} that low-metallicity C-stars 
  have \textit{lower} mass-loss rates, than their more metal-rich counterparts with similar stellar parameters, as long as they 
  have a comparable amount of condensible carbon.}


\maketitle

\section{Introduction}
According to the most commonly accepted scenario, the mass loss of AGB stars is driven by radiation
pressure on dust-grains in the outer stellar atmospheres. This pressure causes the acceleration of the outer layers, which develop, as the
escape-velocity is reached, into a stellar wind. This mechanism is enhanced by the pulsations of these stars and the resulting
shock waves which create conditions for efficient dust condensation. The wind momentum is, however, generated by radiative acceleration, 
since pulsations alone cannot inject sufficient energy to sustain an outflow \citep{Wood79}.

The metallicity of an AGB star can affect its wind properties since it influences the atmospheric structure
and possibly the availability of condensible material for dust formation. Whether low metallicity 
should be associated with smaller mass loss is however unclear. Bowen \& Willson (1991) argued that 
low-metallicity AGB stars should experience a so-called superwind phase, but proposed that this should 
occur at a higher luminosity than for a metal-rich case. Their argument was that at lower 
metallicities the molecular opacities are smaller, which in turn can reduce the extension of the atmosphere. Since the onset 
of dust formation depends on whether the atmosphere has reached a critical size, it appears correct to expect that 
weaker winds occur at lower metallicities. On the other hand, observational studies of metal-poor carbon stars in the Magellanic 
Clouds and the Milky Way indicate similar mass-loss rates \citep{vanLoon00, Groenewegen07}.

We report results of an on going project to explore the mass loss properties of C-rich AGB 
stars as a function of stellar parameters. We question the common idea that AGB stars at lower metallicity should 
have weaker winds, and argue that the injection of kinetic energy by pulsations plays an important r\^ole in wind formation 
and the resultant mass-loss rate. We claim that commonly-made assumptions about boundary conditions in mass-loss models of 
AGB stars make it difficult to compare results in the literature. The simulation of stellar pulsations, using a time-dependent 
inner boundary condition, requires, in particular, a careful choice of parameterisation; for example, results derived using 
different rates of kinetic-energy injection, should not be compared.

Theoretical studies of dust-driven AGB mass loss at sub-solar (LMC) metallicity \citep{Wachter08}, suggested that these 
stars should have slower winds, but that the mass-loss rate should be mostly unaffected by the decrease in metallicity. In
these models, a constant ${\rm C/O}=1.8$ was assumed, which was the same value as in the solar-metallicity models of Wachter 
et al. (2002). Due to this assumption, the LMC-models have a significantly lower abundance of free carbon, i.e., 
carbon that is not bound to CO and may therefore form dust particles. As a consequence, less dust forms. A similar effect 
was observed in the results by Helling et al. (2000), who found that models of LMC metallicity have, in genral, slower winds and a 
lower mass-loss rate. Given the assumptions, these trends are unsurprising, since reducing 
the amount of free carbon, and consequently the dust/gas-ratio, makes momentum transfer from radiation to  
dust and gas less efficient, leading to a slower wind. We note that molecular bands in metal-poor 
C-stars are as strong as, or stronger than the corresponding stars at solar metallicity. This was first observed in the optical 
\citep{Westerlund91} and recently confirmed in the infrared \citep[see, e.g.,][and refereces therein]{Lagadec07}. 
The wind models presented by Wachter et al. (2008) and Helling et al. (2000) are also based on grey radiative transfer, 
which appears to affect the mass-loss rates \citet[cf.][]{Hofner03}. Moreover, the velocity amplitudes and not the rates of energy-injection 
by pulsations were used to quantify the effects of pulsations on mass loss \citep[see, also,][]{Winters00}.

\section{Numerical Modelling}
The models for the atmospheres and winds are computed using frequency dependent radiative transfer for the 
gas and dust, including detailed micro-physics of the dust grains and their formation \citep[described in][]{Andersen03,Hofner03},
in combination with time-dependent hydrodynamics. Here we describe the initial and boundary conditions,
in addition, to further assumptions. 

\subsection{Pulsations and the Inner Boundary Condition}
\begin{figure*}
  \sidecaption
  \includegraphics[width=6cm]{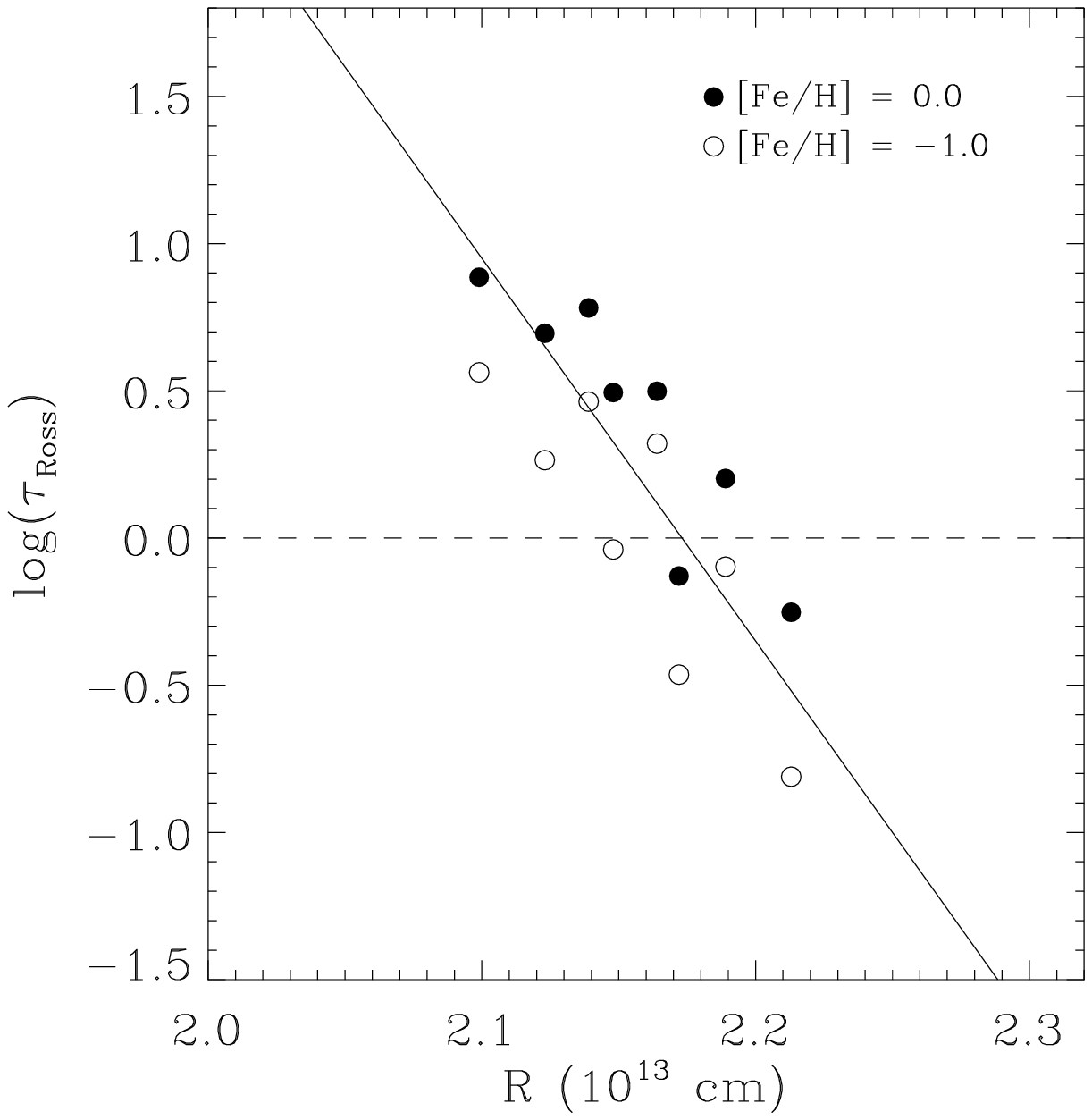}
  \includegraphics[width=6cm]{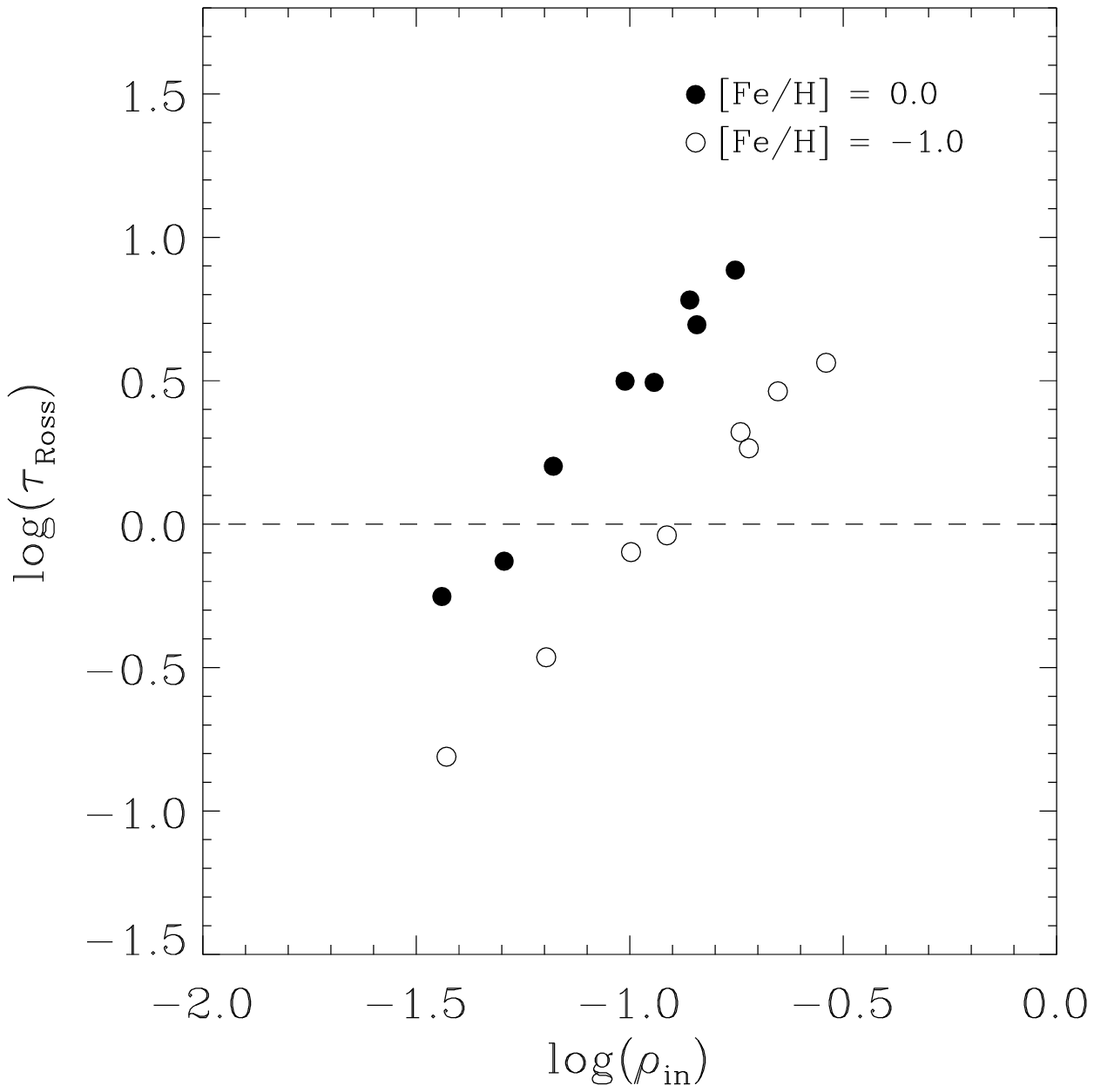}
  \caption{\label{Tau}
  The Rosseland mean opacity at the inner boundary of the hydrostatic initial atmospheric structures as function of the geometric location of 
  the boundary and the gas density at the boundary (at $t = 0$) for solar (Asplund et al. 2005) and a typical metal-poor composition 
  (${\rm [Fe/H]} = -1.0$). The full line in the left panel shows an opacity-law of the form $\tau_{\rm Ross} = \tau_0 e^{-r/R_0+1}$, where
  $\tau_0$ is the opacity at a some radial point $R_0$. The horizontal dashed lines marks $\tau_{\rm Ross} = 1$. Points below this line are 
  usually associated with too shallow initial structures.}
  \end{figure*}

The kappa-mechanism \citep[see, e.g.,][]{Olivier05}, which is believed to be the cause of long-period pulsations, operates in the 
convective envelope. In our models, we simulate the effect of this mechanism on the atmosphere, using a time-dependent inner boundary 
condition, i.e., a so-called piston boundary of the form
\begin{equation}
\label{piston}
R_{\rm in}(t) = R_{\rm in}(0) + {\mathcal{P}\over 2\pi} \Delta u_{\rm p} \sin\left({2\pi\over \mathcal{P}} t\right),
\end{equation}
where $\Delta u_{\rm p}$ is the velocity amplitude and $\mathcal{P}$ is the pulsation period. We employ an empirical period-luminosity
relation \citep{Feast89}, which maintains a direct relationship between the period and luminosity, for all models.
Since the bottom layers of the atmosphere must be optically thick for the model atmosphere to be realistic, one should 
place the piston boundary in that optically thick region of the atmosphere. This leads to inner boundary conditions for 
the metal-poor (SMC-like) models where the gas pressure is significantly higher at $R=R_{\rm in}$ (see Fig. \ref{Tau}), while the temperature 
changes much less. Thus, the kinetic energy-injection by the pulsations/piston is typically larger in these models for a given $\Delta u_{\rm p}$.

\subsection{Free Carbon and Dust Formation}
\label{dust}
The raw-material for dust grains is the condensible carbon in the atmosphere, i.e., the carbon that is not bound in CO molecules and
therefore, in principle, available for dust formation. Since the absolute oxygen abundance $\varepsilon_{\rm O}$ may vary depending on the 
initial metallicity of the star and the amount of oxygen dredge-up, $\tilde{\varepsilon}_{\rm C} \approx \varepsilon_{\rm C} - \varepsilon_{\rm O}$ is
a more relevant quantity than ${\rm C/O}=\varepsilon_{\rm C} / \varepsilon_{\rm O}$. 

Low-metallicity C-stars do not necessarily have carbon-poor atmospheres. Since oxygen (along with large amounts of carbon) is most likely 
dredged-up in each thermal-pulse cycle \citep[see, e.g.,][and references therein]{Herwig05} the ${\rm C/O}$-ratios do not have to be large, 
even if $\tilde{\varepsilon}_{\rm C}$ is relatively large. For C-stars with a solar (or higher) relative abundance of oxygen, the oxygen abundance 
is hardly affected by oxygen dredge-up, but for metal-poor C-stars, where the inital abundance is small, it certainly is. We
emphasise that observations by Wahlin et al. (2006) indicate that ${\rm C/O}$-ratios in metal-poor C-stars may be almost an order 
of magnitude higher than at solar metallicity; so even without oxygen dredge-up, $\tilde{\varepsilon}_{\rm C}$ may therefore be quite large. 
After a number of thermal pulses, there may be as much free carbon in a metal-poor C-star atmosphere as for a solar-metallicity.
 
  \begin{table}
  \caption{\label{models} CNO abundances and metallicity of the models considered in this letter.}
  \begin{tabular}{lccccr}
  \hline
  \hline
   & $\log(\tilde{\varepsilon}_{\rm C})$ & $\log(\varepsilon_{\rm C})$ & $\log(\varepsilon_{\rm N})$ 
        & $\log(\varepsilon_{\rm O})$ & ${\rm [Fe/H]}$ \\
  \hline
  AGS05 & $8.50/8.80$ & $8.89/9.04$ &  $7.78$ & $8.66$ &  $0.0$\\
  Sub-solar  & $8.50/8.80$ & $8.81/8.86$ &  $6.78$ & $7.96$ & $-1.0$\\ 
  \hline
  \end{tabular}
  \label{models}
  \end{table}

We assume that $\tilde{\varepsilon}$ is almost unrelated to the over all initial metallicity of the star. We use the same
$\tilde{\varepsilon}$ for subsolar and solar-metallicity models, i.e., we compare the effects of different metallicities only.
Since $\tilde{\varepsilon}$, and not the ${\rm C/O}$-ratio, remains fixed, it is not obvious that a lower initial metallicity 
leads to lower mass-loss rates, when comparing mass-loss models at different metallicities.

  \begin{table}
  \caption{\label{models2} Parameters used for the models. All models have $M_\star = 1 M_\odot$, 
           A-models have $\log(\tilde{\varepsilon})=8.80$, and B-models $\log(\tilde{\varepsilon})=8.50$.}
  \begin{tabular}{cccccc}
  \hline
  \hline
  Model & $\log(L/L_\odot)$ & $T_{\rm eff}$ (K) & $\log(g)$ & $\mathcal{P}$ (d) & $\Delta u_{\rm p}$ (km s$^{-1}$) \\
  \hline
  A1 & $3.85$ & $2800$ & $-0.65$ & $390$ & $3.0$\\
  A2 & $3.85$ & $2800$ & $-0.65$ & $390$ & $4.0$\\
  A3 & $3.85$ & $2800$ & $-0.65$ & $390$ & $5.0$\\[1mm]
  B1 & $3.70$ & $2600$ & $-0.64$ & $294$ & $4.0$\\
  B2 & $3.70$ & $2600$ & $-0.64$ & $294$ & $5.0$\\
  B3 & $3.70$ & $2600$ & $-0.64$ & $294$ & $6.0$\\
  \hline
  \end{tabular}
  \end{table}
 
\section{Results and Discussion}
\label{results}
\subsection{Numerical Results}
\begin{figure*}
  \includegraphics[width=6cm]{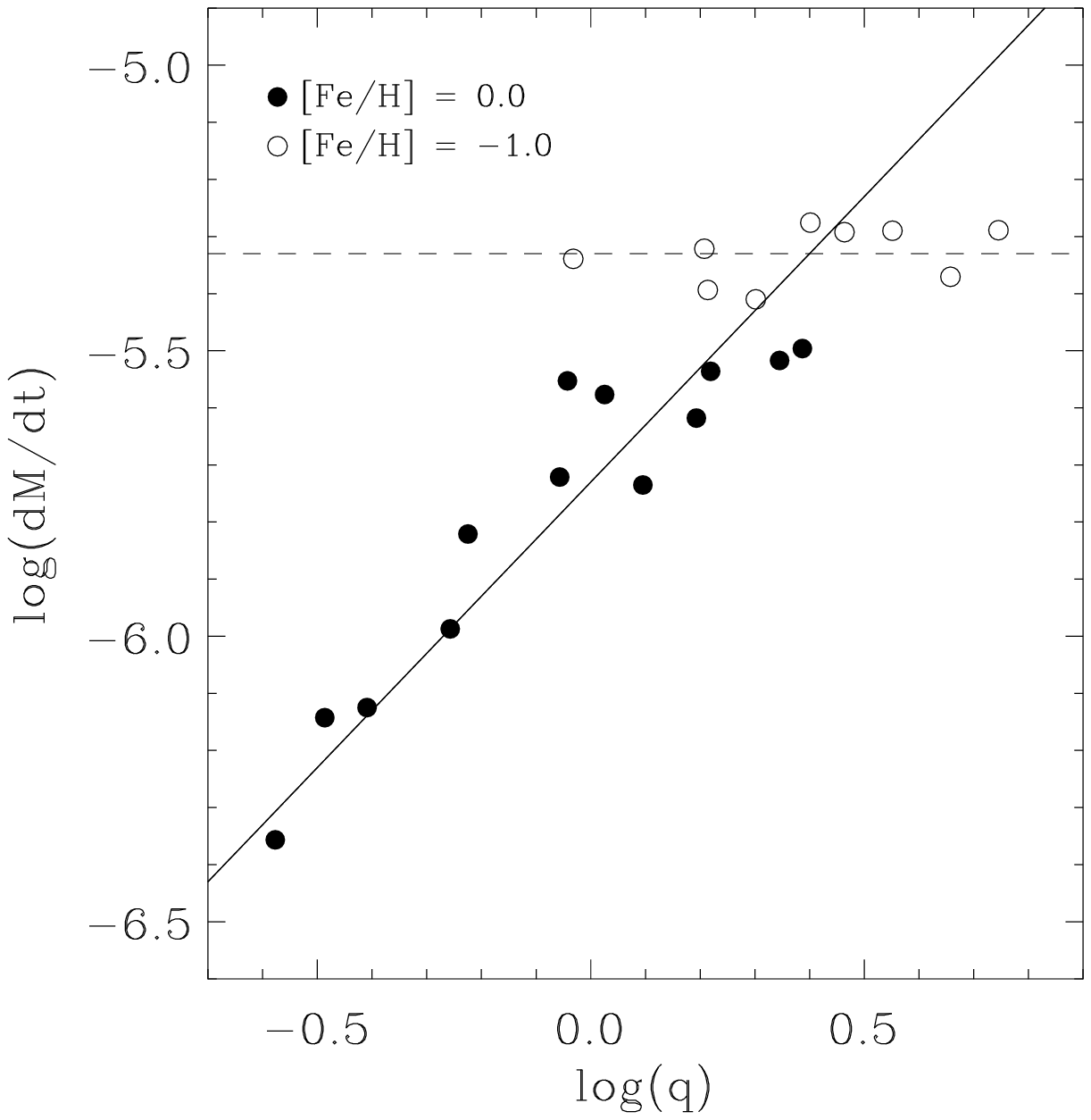}
  \includegraphics[width=6cm]{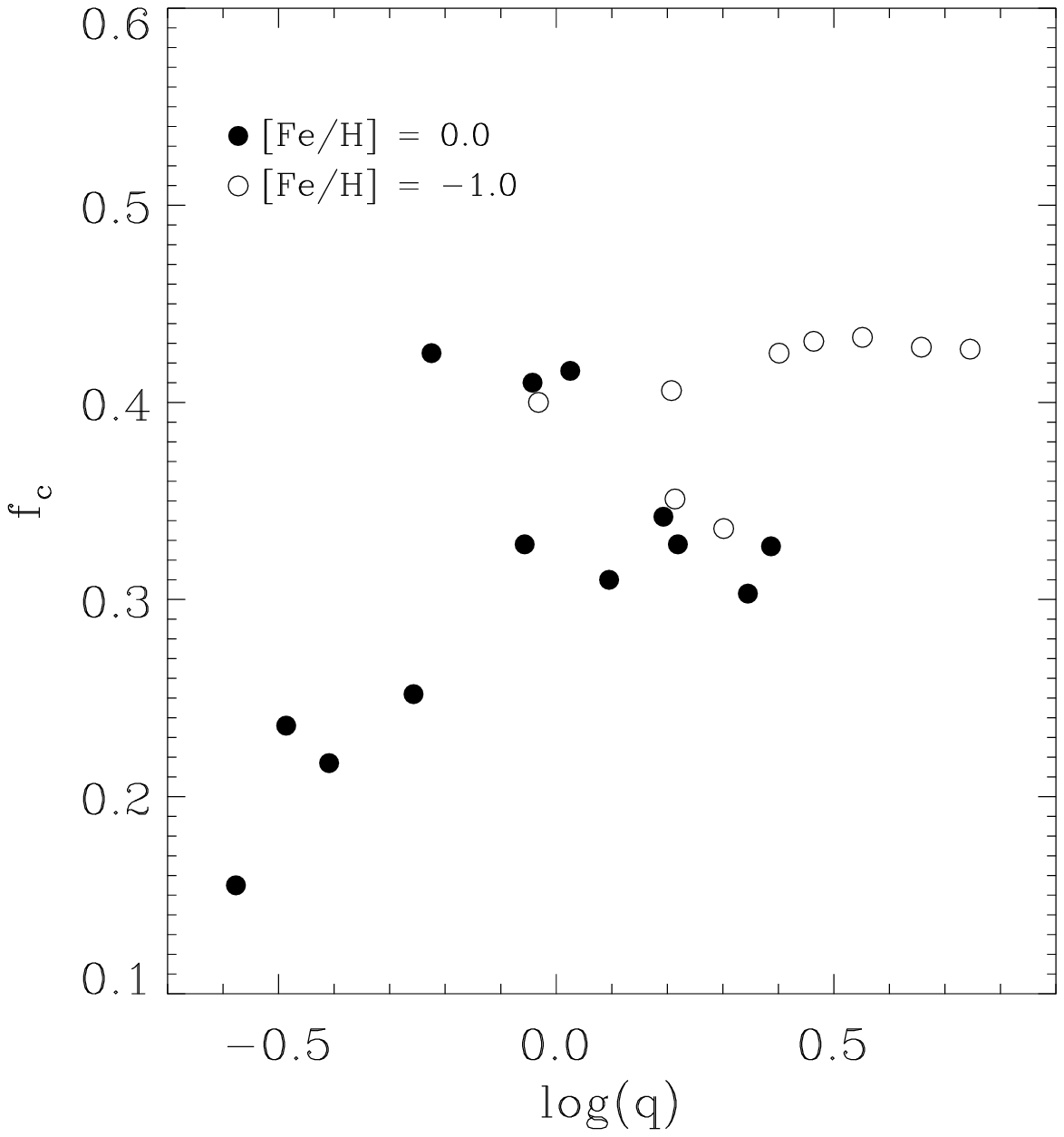}
  \includegraphics[width=6cm]{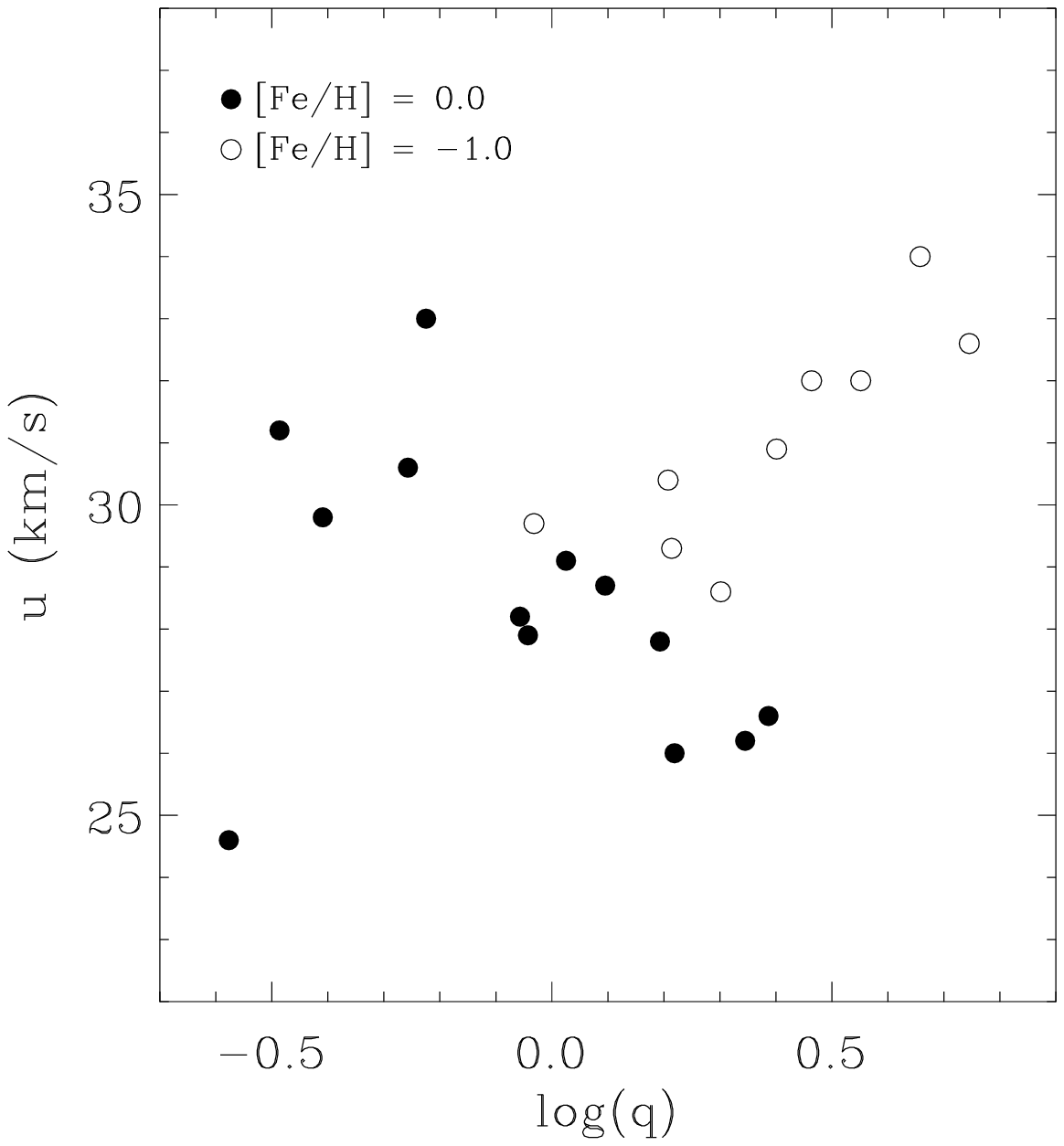}
  \caption{  \label{windprop}
  Comparision between mass-loss rate, mean degree of dust condensation ($f_{\rm c}$), and wind velocity for A-models, where 
  $\log(\tilde{\varepsilon}) = 8.8$ (see also Table \ref{models2}), as functions of the energy-injection 
  parameter $q$ for solar (Asplund et al. 2005) and a typical metal-poor composition (${\rm [Fe/H]} = -1.0$).
  The solid and dashed lines in the left panel indicate $\langle\dot{M}\rangle \propto q$ and $\langle\dot{M}\rangle = {\rm const.}$, respectively.}
  \end{figure*}

\begin{figure*}
  \includegraphics[width=6cm]{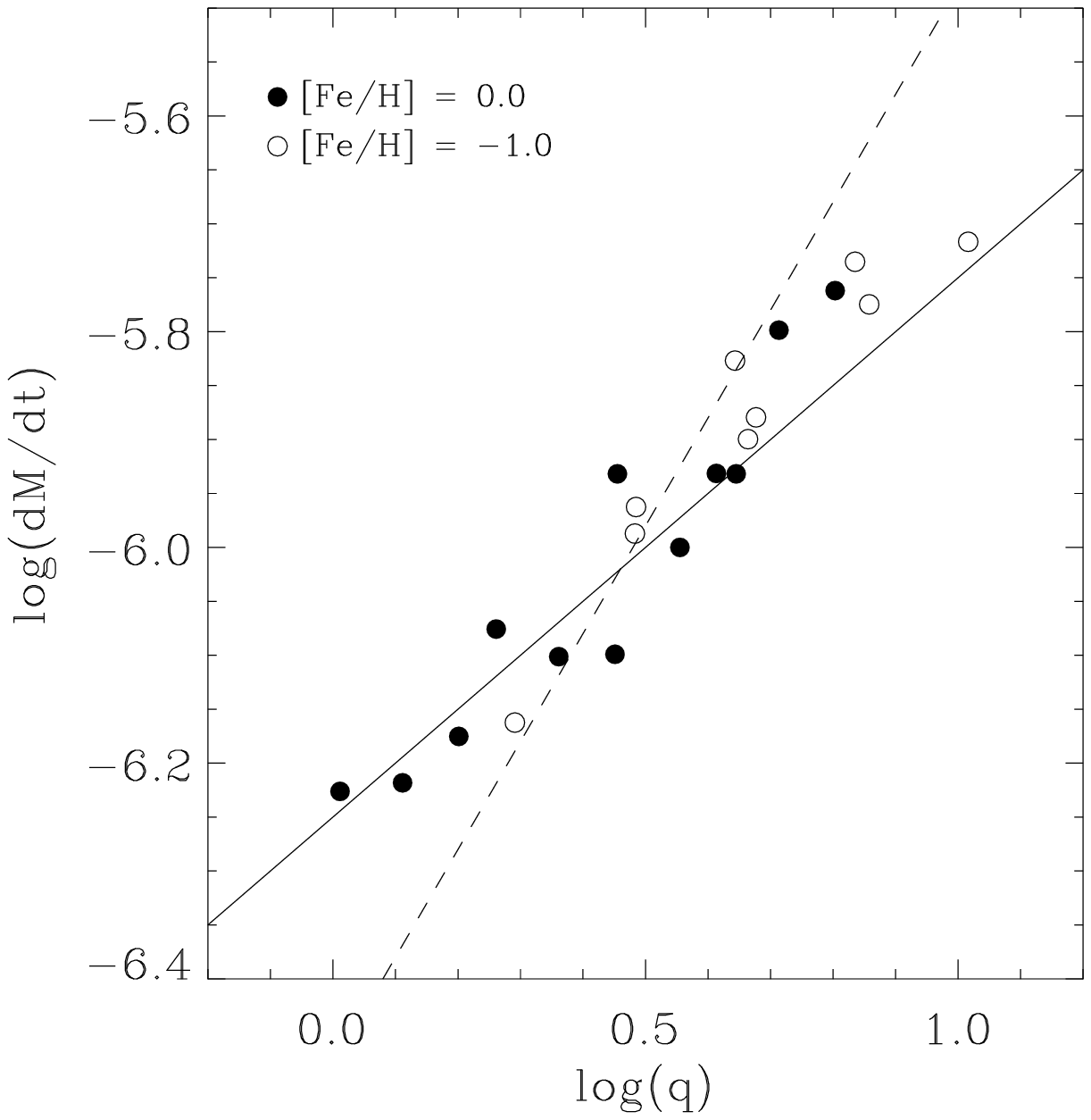}
  \includegraphics[width=6cm]{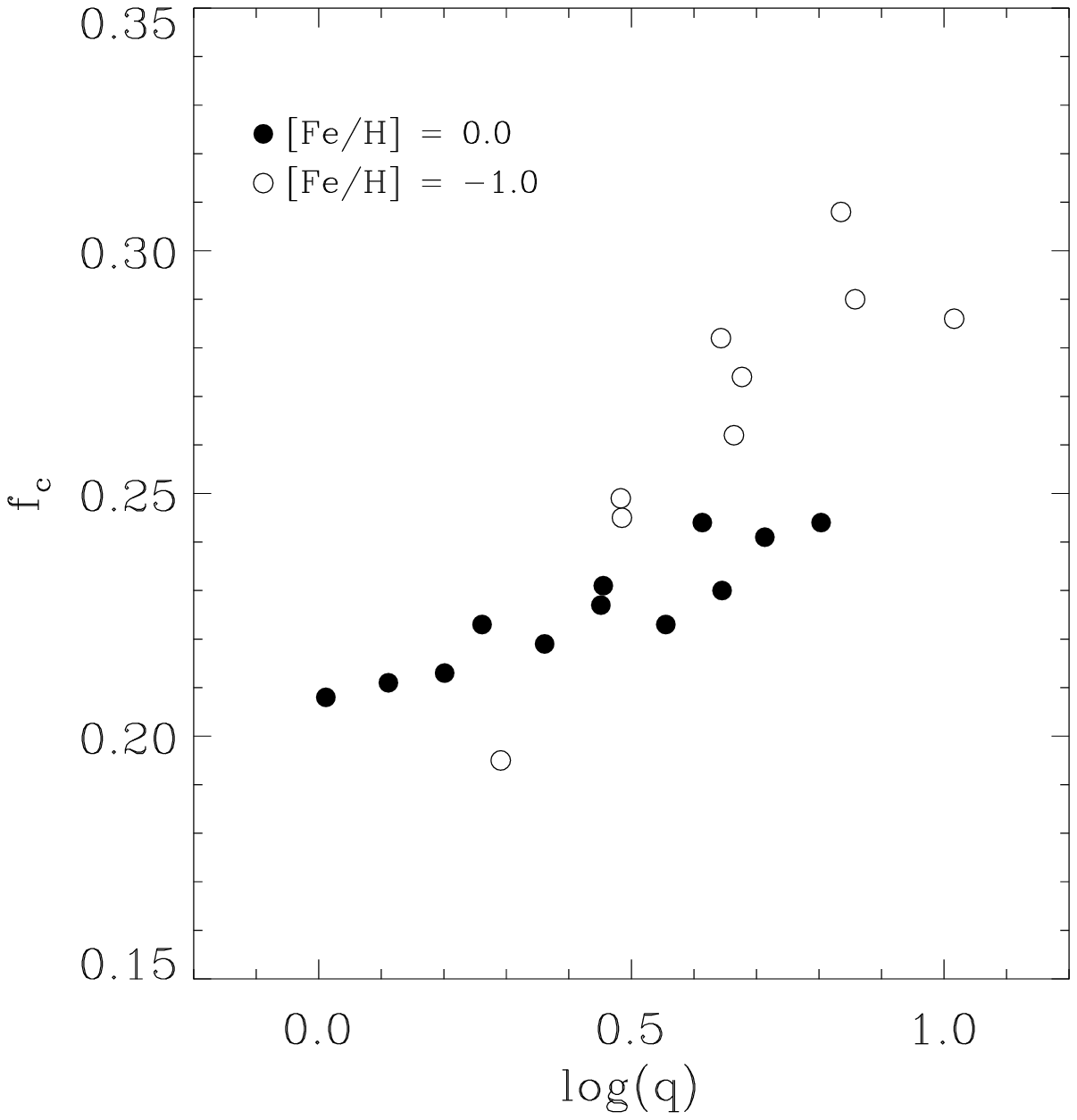}
  \includegraphics[width=6cm]{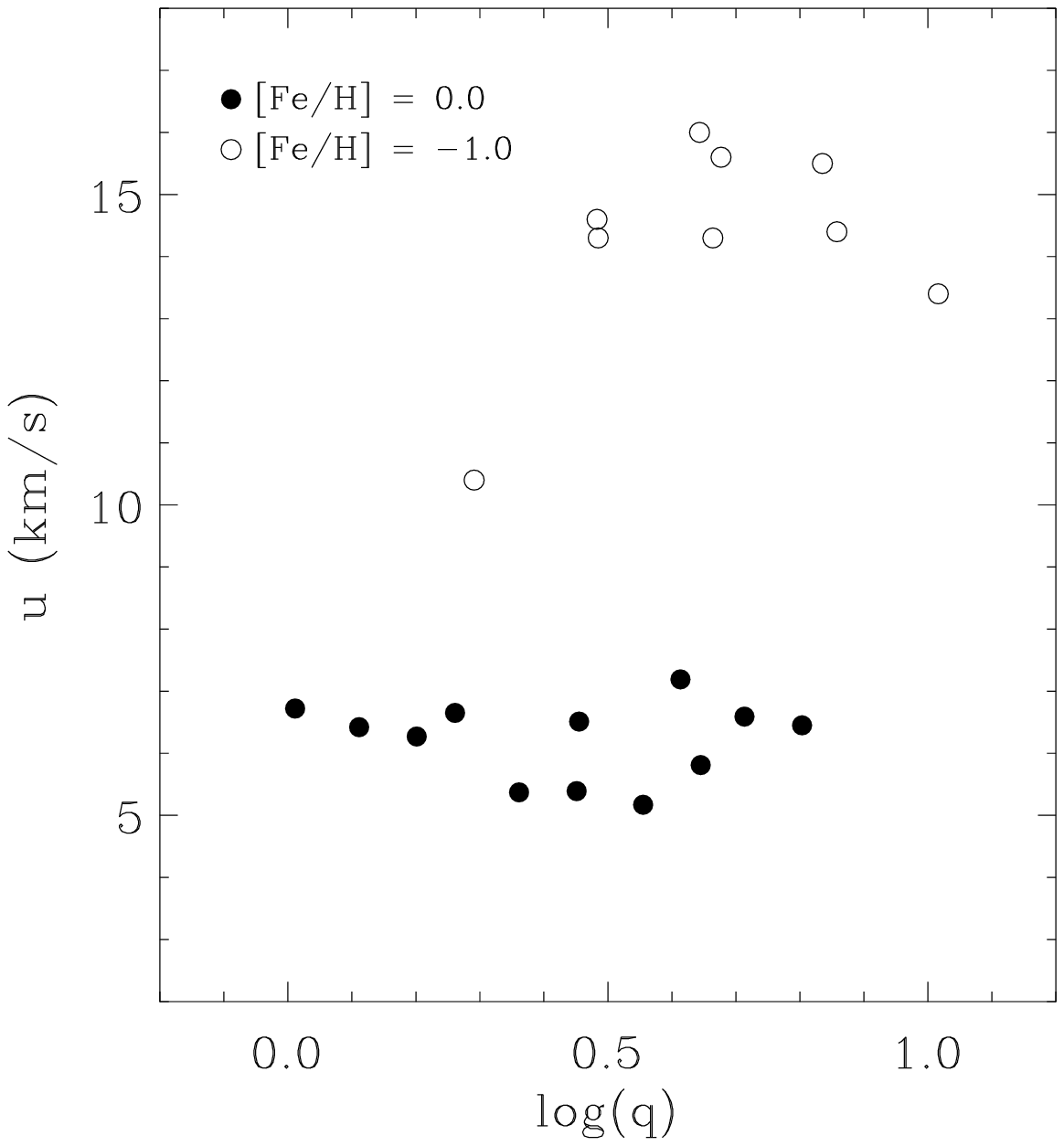}
  \caption{  \label{windprop2}
  Comparision between mass-loss rate, mean degree of dust condensation ($f_{\rm c}$), and wind velocity for B-models, where 
  $\log(\tilde{\varepsilon}) = 8.5$ (see also Table \ref{models2}), as functions of the energy-injection 
  parameter $q$ for solar (Asplund et al. 2005) and a typical metal-poor composition (${\rm [Fe/H]} = -1.0$).
  The solid and dashed lines in the left panel indicate $\langle\dot{M}\rangle \propto q^{1/2}$ and $\langle\dot{M}\rangle \propto q$ , respectively.}
  \end{figure*}

We computed a set of models, where we considered two sets of stellar parameters (labelled A and B), two chemical compositions 
(see Table \ref{models} and Table \ref{models2}), and various positions of the inner (piston) boundary. The results are the following (see also Fig. 
\ref{windprop} and Fig. \ref{windprop2}). For a fixed value of $\tilde{\varepsilon}_{\rm C}$ and a given set of stellar 
parameters, we found that:
\begin{itemize}
\item  The location of the inner boundary can affect significantly the mass-loss rate if $\Delta u_{\rm p}$ is kept fixed.
\item  Models with a metal-poor composition (apart from carbon) have similar, or higher, mass-loss rates compared to models 
       of solar composition (defined as that of Asplund et al. 2005) if the kinetic-energy injections by pulsations are similar.
\item  Models with a solar composition, in general, show slower winds, lower mass-loss rates, and lower degrees of dust condensation 
       than models with metal-poor composition for fixed $\Delta u_{\rm p}$ if the inner boundary
       is located at smiliar optical depth, due to the difference in density at that point.
\end{itemize}  
We emphasise that we compare models with the same absolute abundance of free carbon, which excludes the possibility 
that these effects are due to the accessibility of raw material for dust formation (see Sect. \ref{dust} for further details). 
However, the different absolute abundances of elements other than carbon affect molecular opacities and, consequently,
the structure of the atmosphere. Compared to these abundance differences, the difference in the mass-loss properties 
of both of the models considered (A and B), appears to be more strongly correlated with the input of kinetic energy at the inner boundary.
In our models, we use a sinusoidal piston boundary condition with a velocity amplitude $\Delta u_{\rm p}$, which means that the energy 
injected into a mass shell of mass $dm$ at the inner boundary is $dE_{\rm p}  = {1\over 2} \Delta u_{\rm p}^2\,dm$. 
From an observational point of view, the mass loss rate appears to be proportional to the inferred pulsational energy injection rate of 
AGB stars \citep[see][and references therein]{vanLoon08}. We therefore define a quantity
\begin{equation} 
q \equiv \Delta u_{\rm p}^2\,\rho_{\rm in}(t=0),
\end{equation}
which is proportional to the kinetic-energy injection, and compare with the wind properties. We find that, in general,
\begin{equation}  
\log(\dot{M})\propto \log(q),
\end{equation}
while the wind velocity $u_{\rm out}$ shows a more complicated dependence on $q$ (see Fig. \ref{windprop} and Fig. \ref{windprop2}). 
The latter is caused by the levitation of the outer atmospheric layers and the increase in density, which affects the 
dust-formation efficiency.

In Fig. \ref{windprop} and Fig. \ref{windprop2}, we note that the B-models with metal-poor composition, show typically faster 
winds than solar metallicity B-models, and that metal-poor A-models show mass-loss rates higher than the corresponding models 
with solar composition (see Fig. \ref{windprop}). We note also that the mass-loss rates and mean degrees of dust condensation 
($f_{\rm c}$-values), obtained for the 
A-models with metal-poor composition, in principle, show no trends with $q$; in these cases, this is because the region of wind 
formation is saturated with dust, i.e., $f_{\rm c}$ reaches its upper limit (typically $\sim 0.4-0.5$), and the wind is maximally dust-driven.
Furthermore, the metal-poor A-models with $\Delta u_{\rm p} \la 2$ km s$^{-1}$ cannot sustain a wind due to inefficient 
dust formation. The same is true for B-models with $\Delta u_{\rm p} \la 3$ km s$^{-1}$. This is because the temperature gradient 
is slightly flatter in the AGS05 models, which affects the conditions for dust formation negatively and therefore requires a higher 
kinetic-energy input to trigger dust formation.

\subsection{The $\dot{M}$-$q$ Connection}
From the above, it is obvious that there is a connection between the average mass-loss rate and the $q$-parameter. In principle we find three
different cases: $\dot{M}\propto q^{1/2}$, $\dot{M}\propto q$, and $\dot{M} = {\rm const}$., which can be derived 
from the equation of motion for the atmosphere and wind; a detailed discussion is beyond the scope of this letter and will be published in a
forthcoming paper. Before we discuss the consequences (see Sect. \ref{lowz}), it is of interest to discuss how these dependences on $q$ 
arise from a physical point of view. 

The first two cases reflect the effects of the energy-input by
pulsations, which affects the levitation of the outer atmospheric layers and mean degree of dust condensation. Hence, the mass-loss rate increases
as more kinetic energy is pumped into the atmosphere. The third trend, or the absence of a trend, is due to saturation of the dust-formation 
mechanism, i.e., there is an upper limit for the dust/gas-ratio. When this limit is reached, the dust opacity does not change
when more kinetic energy is pumped into the atmosphere and the mass-loss rate remains constant. The transitions between these cases
depend on the period/luminosity of the model stars, and their atmospheric structures.

\subsection{Stronger Winds at Lower Metallicity?}
\label{lowz}
As is evident in Fig. \ref{windprop} and Fig. \ref{windprop2}, a larger $q$-value 
corresponds, in general, to a larger mass loss and the models with a metal-poor composition have typically faster winds and higher mass-loss 
rates for given stellar parameters. They also require larger $q$-values to sustain a wind. In the case of the A-models, 
the trends (mass-loss rates as a function of $q$) appear to be different for the two metallicities considered, although the number of models 
is too small to establish the existence of two different trends statistically. However, in all cases, our modelling results suggest that the winds 
may become stronger at sub-solar metallicities, or at least not weaker than at solar metallicity, provided that the
abundance of free carbon is the same even if the over all metallicity is different. This result may explain observations 
of mass-loss rates of C-stars in the LMC, and other local dwarf galaxies \citep{vanLoon00, vanLoon03, Matsuura07}, in which the mass-loss rates 
appear to be as high, or even higher, than for C-stars in the solar neighbourhood. Similar results were obtained by Jackson et al.
(2007a; 2007b) for C-stars in other dwarf galaxies.

\section{Summary and Conclusions}
The results presented above may be summarised as follows:\\[2mm] {The mass loss of C-rich AGB-stars, for a given set of stellar
parameters and abundance of free carbon, is a function of the energy-injection by pulsations as long as the dust formation has not
reached the saturation limit. When this limit is reached, the mass-loss rate no longer increases with the energy-injection.}\\[2mm] 

The range of realistic, or even possible, pulsation energies is limited, and connected to the stellar parameters. The 
treatment of the inner boundary and pulsations in the case of a non-saturated wind is crucial. Hence, the importance 
of constraints on the energy injection by pulsations cannot be emphasised enough. A quantitatively-correct theory, or reliable 
observational constraints, of long-period stellar pulsation are required to obtain a complete quantitative understanding of mass 
loss on the AGB. Since the results here are based on only a rather small grid of models, further study of 
boundary-condition effects is needed and will be presented in a forthcoming paper. 

However, one firm conclusion can be drawn from present results: {\it low metallicity should not necesserily be associated with a lower mass-loss 
rate}. More precisely, the development of a strong super-wind phase is expected for C-stars at low (sub-solar) metallicity. The duration of the 
AGB-phase may, in particular, be affected and the effects on nucleosynthesis results (stellar yields) may be profound.

\begin{acknowledgements} 
The authors wish to thank the referee, J.Th. van Loon, for his
careful reading of the manuscript and constructive criticism that helped to improve
the clarity of this paper. B. Gustafsson is thanked for his critical reading and valuable comments on the manuscript. 
A. Wachter is thanked for making unpublished results on mass loss at low metallicity available to us.
\end{acknowledgements}

\bibliographystyle{aa}

\end{document}